\documentstyle[12pt,psfig,epsf]{article}
\newcommand{\be}{\begin{equation}}
\newcommand{\qee}{\end{equation}}
\newcommand{\bea}{\begin{eqnarray}}
\newcommand{\eea}{\end{eqnarray}}

\begin{document}

\hfill  NRCPS-HE-2001-4

\vspace{24pt}
\begin{center}
{\large \bf The Spectrum of the Loop Transfer Matrix on Finite Lattice
}

\vspace{24pt}
{\sl Georgios Daskalakis and George K.Savvidy}

National Research Center Demokritos,\\
Ag. Paraskevi, GR-15310 Athens, Hellenic Republic 

\end{center}
\vspace{12pt}

\centerline{{\bf Abstract}}

\vspace{23pt}
\noindent
We consider the model of random surfaces with extrinsic curvature term 
embedded into 3d Euclidean lattice $Z^3$. On a 3d Euclidean lattice it 
has equivalent representation in terms of transfer matrix 
$K(Q_{i},Q_{f})$, which describes the propagation of loops $Q$.
We study the spectrum of the transfer matrix $K(Q_{i},Q_{f})$ on finite 
dimensional lattices.  The renormalisation group technique is used 
to investigate phase structure of the model and its critical behaviour.


\newpage

\pagestyle{plain}
\section{Introduction}

Various  models of  random surfaces built out of 
triangles embedded into continuous space $R^d$ 
and surfaces built out of  plaquettes  
embedded into Euclidean  lattice $Z^d$ have been considered in the literature 
\cite{weingarten,gross}.
These models  are based on area action and suffer the problem 
of non-scaling behaviour of the string tension and the dominance of 
branched polymers \cite{ambjorn}.   
Several studies have analyzed the physical effects produced 
by rigidity of the surface introduced by adding 
dimension-less extrinsic curvature term to the area action \cite{helfrich}.
Comprehensive review of the work in this area up to 1997 can be 
found in \cite{ambjorn}. 

In this article we shall consider a  model of random 
surfaces $solely$ based on the with extrinsic curvature term  embedded 
into 3d Euclidean lattice $Z^3$  \cite{amb}. 
The corresponding transfer matrix has the form \cite{sav2}
\be
K(Q_{1},Q_{2}) = 
\exp \{-\beta ~[ k(Q_{1}) + 
2 l(Q_{1} \bigtriangleup Q_{2}) + k(Q_{2})]~ \}, \label{tranmat}
\qee
where $Q_{1}$ and $Q_{2}$ are closed polygon-loops on a two-dimensional 
lattice, $k(Q)$ is the curvature and $l(Q)$ is
the length of the polygon-loop $Q$ \footnote{We shall use the 
word "loop" for the "polygon-loop".}. 
This transfer matrix describes  the propagation 
of the initial loop $Q_{1}$ to the final loop $Q_{2}$. 

The spectrum of the transfer matrix which depends 
only on symmetric difference of initial and final loops
$Q_{1} \bigtriangleup  Q_{2}$ 
\be
\tilde{K}(Q_{1},Q_{2}) = 
\exp \{-2 \beta  l(Q_{1} \bigtriangleup Q_{2}) ~ \}, \label{subtranmat}
\qee
has been evaluated analytically in terms 
of correlation functions of the 2d Ising model in \cite{george}.

Our aim in this article is to consider the full transfer 
matrix (\ref{tranmat})   which depends not only on 
symmetric difference of loops $Q_{1} \bigtriangleup  Q_{2}$,  
but also on individual curvature terms $k(Q_{1})+ k(Q_{2})$.
Because analytical solution of this problem is not known yet, 
we shall consider numerical evaluation of the transfer matrix 
on finite lattices in order to gain new insight into spectral 
properties of the transfer matrix in general case and critical 
properties of the model.

In the next sections we shall numerically compute the spectrum of transfer 
matrix on finite Euclidean lattices $T^2$ of the sizes 
$N \star M$, where $N,M=1,2,3,4$. Number of loops on these lattices
$\gamma = 2^{N M}$ will be  $2,4,16,64,512,4096$, $65536$
and transfer matrices would be of the size 
$\gamma \times \gamma  =2^{N M} \times 2^{N M}$.

\section{Basic Formulae}

The partition function is defined as \cite{sav2}
\be
Z(\beta) = 
\sum_{\{Q_{1},Q_{2},...,Q_{\tau}\}}~ K_{\beta}(Q_{1},Q_{2})\cdots 
K_{\beta}(Q_{\tau},Q_{1})  = tr K^{\tau}_{\beta}\label{trace},
\qee
where $K_{\beta}(Q_{1},Q_{2})$ is the transfer matrix of  size 
$\gamma \times \gamma $, defined above (\ref{tranmat}).  
$Q_{1}$ and $Q_{2}$ are closed polygon-loops on a two-dimensional 
lattice 
$T^{2}$ of size $N \times M$ and $\gamma = 2^{N M}$ is the total 
number of polygon-loops 
on a lattice $T^{2}$. 
The transfer matrix (\ref{tranmat}) can be viewed as 
describing  the propagation 
amplitude of the polygon-loop $Q_{1}$ at time $\tau$ to 
another polygon-loop $Q_{2}$ 
at the time $\tau +1$. 

The functional $k(Q)$ is the total curvature of the polygon-loop $Q$ which is 
equal to the number of corners of the polygon ( the vertices with 
self-intersection are not counted!) and $l(Q)$ is the length of $Q$ 
which is equal to the number of its links.  The  length functional 
$l(Q_{1} \bigtriangleup  Q_{2})$ is defined as \cite{sav2}
\be
l(Q_{i})+l(Q_{i+1})-2 \cdot l(Q_{i} \cap Q_{i+1} ) 
= l(Q_{i} \bigtriangleup  Q_{i+1}) , \label{common}
\qee
where the polygon-loop~ $Q_{1} \bigtriangleup  Q_{2}~ \equiv~ Q_{1} \cup  
Q_{2} ~\backslash~ Q_{1} \cap  Q_{2} $~ is a union of links 
$Q_{1} \cup Q_{2}$ without common links $Q_{1} \cap  Q_{2}$.
The operation
$\bigtriangleup$ maps two polygon-loops $Q_{1}$ and $Q_{2}$
into a  polygon-loop $Q= Q_{1} \bigtriangleup  Q_{2}$.

The eigenvalues of the transfer matrix $K_{\beta}(Q_{1},Q_{2})$ define all 
statistical properties of the system and can be found as a solution of the 
following integral equation in the loop space 
$\Pi$ \cite{george}
\be
\sum_{\{Q_{2}\} }K_{\beta}(Q_{1},Q_{2})~\Psi(Q_{2})=
\Lambda(\beta)~\Psi(Q_{1})  \label{inte},
\qee
where $\Psi(Q)$ is a function on the loop space. 
The eigenvalues define the partition function (\ref{trace})
\be
Z^{3d}_{NM}(\beta)= \Lambda^{\tau}_{0}(\beta,NM)+...
+\Lambda^{\tau}_{\gamma -1}(\beta,NM) 
 \label{limit},
\qee
and in the thermodynamical limit the free energy is equal to 
\be
-\beta~ f_{3d}(\beta) = \lim_{\tau N M \rightarrow \infty} 
\frac{1}{\tau N M}~ln~ Z^{3d}(\beta)  =  
\lim_{N,M \rightarrow \infty}\frac{1}{N M}~ln~ \Lambda_0(\beta,MN) ,
\qee
where $\Lambda_0(\beta,NM)$ is the largest eigenvalue defined 
at temperature $\beta$ for the lattice of the size $N\times M$. 
The correlation length is defined by the ratio of eigenvalues 
$\Lambda_{1}(\beta,NM)$ and $\Lambda_{0}(\beta,NM)$ as 
$$
\xi_{NM}(\beta) = -1/ln[\Lambda_{1}(\beta,NM)/\Lambda_{0}(\beta,NM)],
$$
and grows if the eigenvalue $\Lambda_{1}$ approaches to the eigenvalue 
$\Lambda_{0}$ at some critical temperature $\beta_{c}$. By  
the Frobenius-Perron theorem $\Lambda_{0}(\beta)$ is simple and  
$
\Lambda_{0}(\beta) > \Lambda_{1}(\beta) \geq \Lambda_{2}(\beta) \geq ...
$

Because we can analyze numerically only finite-dimensional matrices we shall use 
renormalisation group approach to critical phenomena \cite{nigh} in order to 
extract critical indexes. The renormalisation group equation 
\be
\bar{\beta} = \bar{\beta}(\beta) \label{reeq}
\qee
is  defined through the correlation function
\be
{\xi_{NM}(\beta) \over N + M} =    
{\xi_{ \bar{N}\bar{M} }(\bar{\beta}) \over \bar{N} + \bar{M}} ,~~~~~N+M > \bar{N}+\bar{M}.
\qee
The critical point $\beta_{c}$ is a fixed point $\beta^*$ of the renormalisation 
group equation (\ref{reeq})
\be
\beta^{*} = \bar{\beta}(\beta^{*}). \label{crreeq}
\qee
The renormalisation group equation (\ref{reeq}) can be expanded near 
fixed point $\beta^*$ into the series of the following form
$$
\bar{\beta}(\beta) - \beta^* = g_{1}(\beta - \beta^*) + g_{2}(\beta - \beta^*)^2 +...
$$ 
and the critical index $\nu$ is equal to 
\be
\nu = {ln [(N+M)/(\bar{N} + \bar{M} )] \over lng_1}.
\qee
We shall use these formulas to define fixed point and the critical indexes. 

In the next sections we shall explicitly construct transfer 
matrices on finite Euclidean lattices $T^2$ of the sizes 
$N \star M$, where $N,M=1,2,3,4$. Number of loops on these lattices
$\gamma = 2^{N M}$ will be  $2,4,16,64,512,4096$, $65536$
and transfer matrices would be of the size 
$\gamma \times \gamma  =2^{N M} \times 2^{N M}$.

\section{General properties of the spectrum}

We shall fix the boundary spins on a lattice $T^2$, let us say, in up direction. 
The convention is that we shall not count the boundary spins, thus if 
the lattice has the size $2 \times 3$ then it has six internal spins 
\footnote{And ten boundary spins which we shall not count.}. In this setting
when all spins are in up direction then in the dual picture we 
shall have just an empty loop. When one spin inside lattice is turned down, then in the 
dual picture we shall have simply one loop consisting of four bonds, a box-loop, and so on. 

Let us begin from the simplest lattice of the size $1 \times 1$ and then move to 
more complicated ones in order to understand general properties of the spectrum.
For this lattice which has just one spin we have empty loop $\emptyset$ and 
one box-loop $\Box$. The transfer matrices $K$ and $\tilde{K}$ describing 
transitions between these two loops can be computed by using definition (\ref{tranmat}) 
and (\ref{subtranmat})
\be
K = \left( \begin{array}{lr}
 1& e^{-12\beta}\\
 e^{-12\beta}&e^{-8\beta} 
 \end{array} \right),~~~~~~~~ \tilde{K} = \left( \begin{array}{lr}
 1& e^{-8\beta}\\
 e^{-8\beta}&1 
 \end{array} \right), \label{two}
\qee
and then it is easy to find its eigenvalues: 
\be
2 \Lambda_{0,1} = 1+e^{-8\beta} \pm \sqrt{1 + 4e^{-24\beta} + e^{-16\beta} - 2 e^{-8\beta} },~~~~~
\tilde{\Lambda}_{0,1} = 1 \pm e^{-8\beta}.
\qee
\begin{figure}
\centerline{\hbox{\psfig{figure=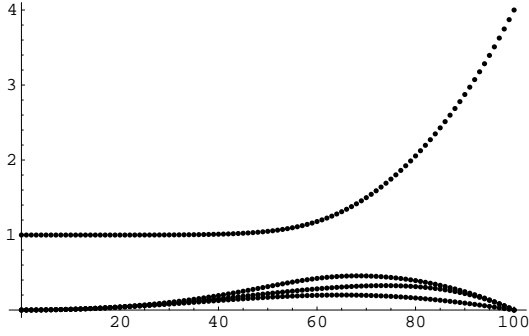,width=7cm}}}
\caption[fig1]{ The typical behaviour of the eigenvalues. These are 
eigenvalues of the transfer matrix K as a function of  $\omega= e^{-4\beta}$ for the 
size $4\times 4$. 
}
\label{fig1}
\end{figure}
For the lattice of the size $1 \times 2$ we have four loops: the empty loop, 
two one-box loops and one two-box loop. The transfer matrices are therefore 
four by four:
\be
K = \left( \begin{array}{llll}
 1& e^{-12\beta}&e^{-12\beta}&e^{-16\beta}\\
 e^{-12\beta}&e^{-8\beta}&e^{-20\beta}&e^{-16\beta}\\
 e^{-12\beta}&e^{-20\beta}&e^{-8\beta}&e^{-16\beta}\\
 e^{-16\beta}&e^{-16\beta}&e^{-16\beta}&e^{-8\beta}
 \end{array} \right),~~~~ 
 \tilde{K} =  \left( \begin{array}{llll}
  1& e^{-8\beta}&e^{-8\beta}&e^{-12\beta}\\
 e^{-8\beta}&1&e^{-12\beta}&e^{-8\beta}\\
 e^{-8\beta}&e^{-12\beta}&1&e^{-8\beta}\\
 e^{-12\beta}&e^{-8\beta}&e^{-8\beta}&1
 \end{array} \right), \label{four}
\qee
The eigenvalues of the  matrix $\tilde{K}$ have been found in \cite{george}
\be
\tilde{\Lambda}_{0} = 1 + 2 e^{-8\beta} + e^{-12\beta},~~~ 
\tilde{\Lambda}_{1,2} = 1 - e^{-12\beta},~~~ 
\tilde{\Lambda}_{3} = 1 - 2 e^{-8\beta} + e^{-12\beta}, 
\qee
and for the  matrix $K$ we have found eigenvalues on a computer (see Figure \ref{fig1}).
In both cases all eigenvalues are positive. 
\begin{figure}
\centerline{\hbox{\psfig{figure=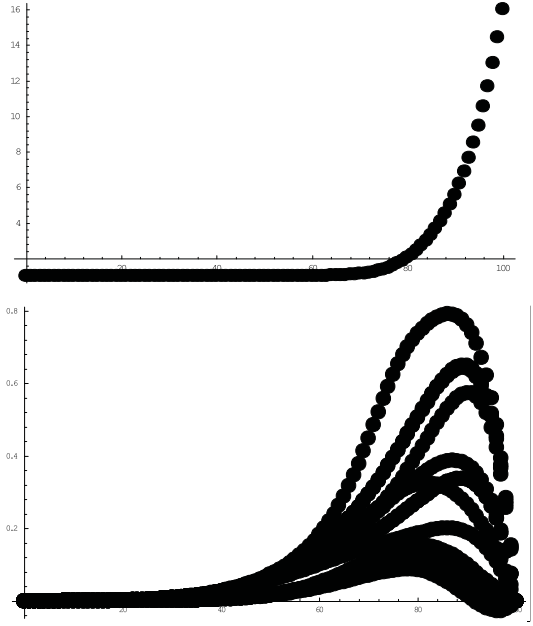,width=7cm}}}
\caption[fig2]{The spectrum of the transfer matrices $K$ of the size 
$16\times 16$ as a function of  $\omega= e^{-2\beta}$.
The qualitative character of the spectrum is the same as in Figure 1, 
but now we can see that at high temperatures some of the 
eigenvalues become negative.}
\label{fig2}
\end{figure}
As the last example of explicitly depicted transfer matrix we shall 
present it for the lattice $2 \times 2$. It has sixteen loops and is of the 
size $16 \times 16$
\be
K = \left( \begin{array}{llllllllllllllll}
 0&6&6&6&6&8&8&8&8&11&11&11&11&11&11&10\\
 6&4&10&12&10&8&12&8&12&9&13&11&13&11&13&12\\
 6&10&4&10&12&8&12&12&8&13&9&13&11&13&11&12\\
 6&12&10&4&10&12&8&12&8&9&13&11&13&11&13&12\\
 6&10&12&10&4&12&8&8&12&13&9&13&11&13&11&12\\
 8&8&8&12&12&4&12&12&12&11&11&9&13&13&9&10\\
 8&12&12&8&8&12&4&12&12&11&11&13&9&9&13&10\\
 8&8&12&12&8&12&12&4&12&9&11&13&13&9&9&10\\
 8&12&8&8&12&12&12&12&4&11&11&9&9&13&13&10\\
 11&9&13&9&13&11&11&9&11&6&14&10&14&10&14&9\\
 11&13&9&13&9&11&11&11&11&14&6&14&10&14&10&9\\
 11&11&13&11&13&9&13&13&9&10&14&6&12&14&12&9\\
 11&13&11&13&11&13&9&13&9&14&10&12&6&12&14&9\\
 11&11&13&11&13&13&9&9&13&10&14&14&12&6&12&9\\
 11&13&11&13&11&9&13&9&13&14&10&12&14&12&6&9\\
 10&12&12&12&12&10&10&10&10&9&9&9&9&9&9&4 
  \end{array} \right), \label{sixteen}
\qee
where we show only the exponents. The eigenvalues of the  matrix $\tilde{K}$ 
have been found in \cite{george}
\be
   \begin{array}{lll}
\tilde{\Lambda}_{0} = 1 + 4e^{-8\beta}+6e^{-12\beta}+ 5e^{-16\beta},& 
\tilde{\Lambda}_{1,2,3,4} = 1 + 2 e^{-8\beta} -3 e^{-16\beta},\\
\tilde{\Lambda}_{5,6,7,8} = 1 - 2e^{-12\beta}+2e^{-16\beta},&
\tilde{\Lambda}_{9,10} = 1 - 2 e^{-12\beta} + e^{-16\beta},\\
\tilde{\Lambda}_{11,12,13,14} = 1 - 2 e^{-8\beta} + e^{-16\beta},&
\tilde{\Lambda}_{15} = 1 - 4 e^{-8\beta} +6 e^{-12\beta} -3e^{-16\beta}.
   \end{array}
\qee
The eigenvalues of matrix $K$  are shown on Figure \ref{fig2}. The general
properties of the spectrum at low and high temperatures can be easily 
understood. At $\beta \rightarrow 0$  the eigenvalues are :
$$
\Lambda_0 = 2^{NM},~~ \Lambda_1 = .....= \Lambda_{2^{NM}-1} = 0
$$
and when $\beta \rightarrow \infty$ they are :
$$
\Lambda_0 = 1,~~ \Lambda_1 = .....= \Lambda_{2^{NM}-1} = 0.
$$
These properties of the eigenvalues at the ends of the 
spectrum can be seen on Figures 1 and 2. These Figures also
allow to see the behaviour of the spectrum at the intermediary 
temperatures, the largest eigenvalue $\Lambda_0$ and the 
second one $\Lambda_1$ approach to each other. The rest 
of the eigenvalues follow the behaviour of the $\Lambda_1$.
As we will see minimal distance between eigenvalue $\Lambda_0$ and 
$\Lambda_1$ decreases when the size of the lattice increases 
signaling the possibility of a phase transition. 

\section{Numerical calculation of eigenvalues $\Lambda_0$ and $\Lambda_1$ }
\begin{figure}
\centerline{\hbox{\psfig{figure=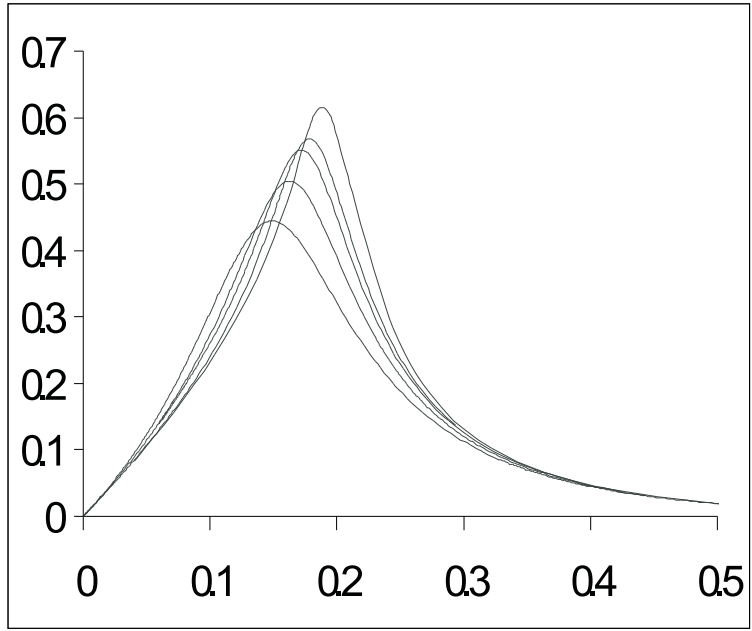,width=10cm}}}
\caption[fig3]{ The ratio  $\Lambda_1/\Lambda_0$ as a function of $\beta$ for the 
matrices of sizes $16\times16, ~64\times 64,~ 256\times 256 ,~512\times512,
 ~4096\times4096$.}
\label{fig3}
\end{figure}

To generate the transfer matrices of the larger size we have developed 
a special fast algorithm to create polygon-loops $Q$ and then to 
compute the elements of the transfer matrix $K$. This part of the 
program is working fast enough to generate matrices of the size 
$65536 \time 65536 $ and higher. To compute  eigenvalues $\Lambda_0$ and $\Lambda_1$ 
we have used standard technique, that is acting by the matrix $K$ on 
an arbitrary vector $\Psi(Q)$ many, $m$, times. In the limit $m \rightarrow \infty$
it converges to $\Psi_{0}(Q)$. In the same manner, acting on the orthogonal 
vector, one can get $\Psi_{1}(Q)$ and therefore corresponding eigenvalues.
The real problem to handle higher size matrices is mostly connected 
with the memory of the computer. 

The ratio of the eigenvalues  $\Lambda_0/\Lambda_1$  is  shown on Figure \ref{fig3}.
This ratio increases with the size of the matrix $K$ and reaches the value $0.6$ for the 
matrix of the size $4096\times4096$. To find fixed point $\beta^*$ of the renormalization 
group equation we have constructed the curves $\xi_{NM}(\beta)/(N+M)$ 
for different lattices $N*M$. As one can see 
on Figure 4 there are two fixed points at $\beta^{*}_{1} \simeq 0.17$ and at
$\beta^{*}_{2} \simeq 0.21$. At the stable fixed point $\beta^{*}_{2} \simeq 0.21$ 
we have computed the value $g_1 \simeq 1.99$ and the critical index $\nu \simeq 0.59$.
This should be compared with the results obtained in \cite{pie} by low 
temperature expansion and by Monte-Carlo 
simulation of the corresponding system in \cite{bathas,cappi}. Our value is 
of the same order of magnitude. 
To confirm these results one should go to high-dimensional matrices.

\begin{figure}
\centerline{\hbox{\psfig{figure=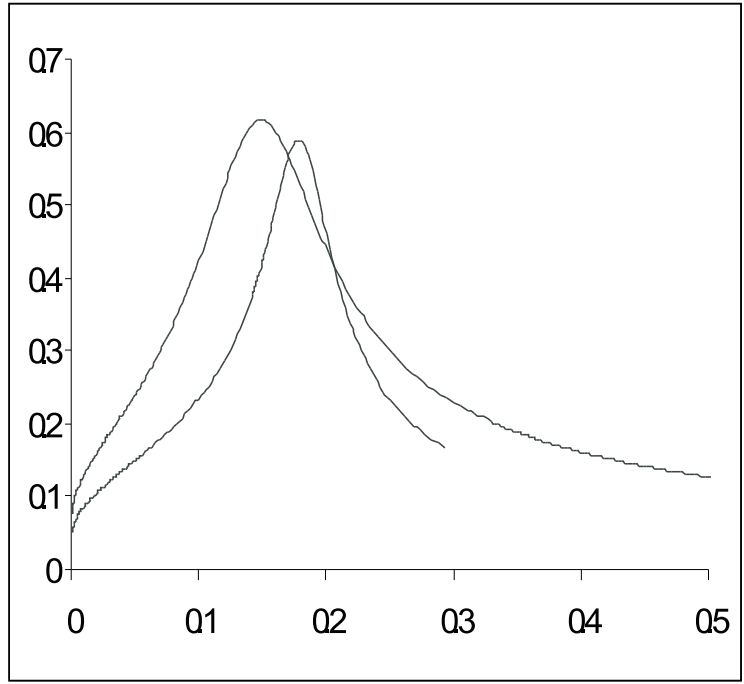,width=10cm}}}
\caption[fig4]{ The ratio  $\xi_{NM}(\beta)/(N+M)$ as a function of $\beta$ for the 
matrices of sizes $16\times16$ and $512\times512$. There are two fixed points at 
$\beta^{*}_{1} \simeq 0.17$ and at $\beta^{*}_{2} \simeq 0.21$.}
\label{fig4}
\end{figure}

It is also interesting to see the behaviour of eigenfunctions. The eigenfunctions $\Psi(Q)$
for the matrix $\tilde{K}(\beta)$ are already known and are independent of the temperature
\cite{george}. This is  because  
transfer matrices $\tilde{K}(\beta)$ commute with each other at different temperatures
and the system is integrable \cite{george}. This is a general property of integrable 
systems \cite{kramers,baxter,zamolodchikov,polyakov}.  
The situation with the full transfer matrices $K(\beta)$ 
is not known, therefore we should check if the eigenfunctions are temperature 
dependent.  For that reason we shall compute the scalar product 
$$
(~ \Psi_{\Lambda}(Q)(\beta') \star \Psi_{\Lambda}(Q)(\beta) ~)
$$
between eigenfunction 
$\Psi_{\Lambda}(Q)(\beta)$ at different temperatures $\beta'$ and  $\beta$ in order 
to see if the scalar product depends on 
temperature. For the first eigenvalue $\Lambda_0$ and $\beta' =0$ the result of the 
calculations is shown on Figure 5. 
It is clearly seen that the scalar product is temperature dependent 
and that it drastically  changes near the critical points. This result 
drives us to the conclusion that full transfer matrix represents
a non-integrable system.

\begin{figure}
\centerline{\hbox{\psfig{figure=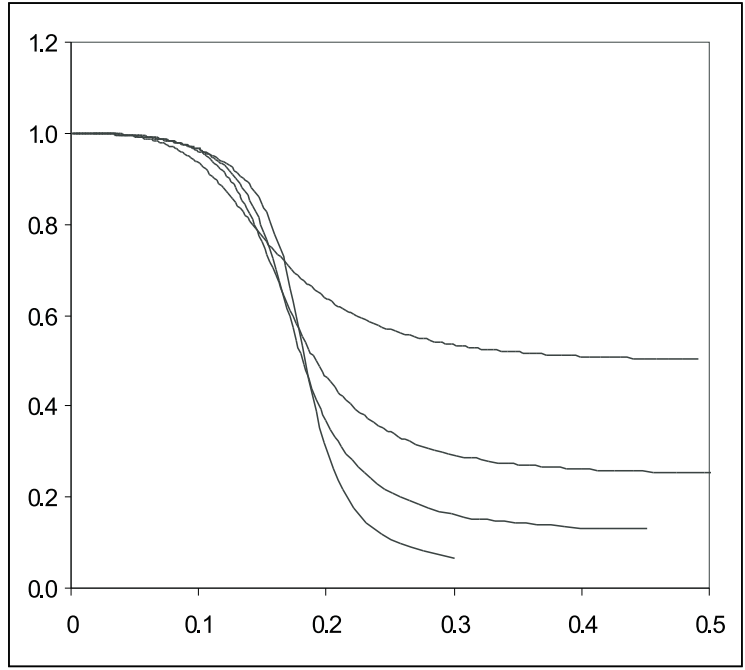,width=10cm}}}
\caption[fig5]{ Temperature dependents of the scalar product  
$\{ \Psi_{0}(Q)(0) \star \Psi_{0}(Q)(\beta) \}$ 
 for the matrices of the  sizes  $16\times16, ~64\times16 ,~512\times512,
 ~4096\times4096$. 
}
\label{fig5}
\end{figure}

We would like to thank Thordur Jonsson for stimulating discussions and 
A.Peppas for his collaboration in early 
stages of this work and calculation of eigenvalues for low dimensional 
matrices. This work was supported in part by the EEC Grant no. HPRN-CT-1999-00161
and the Grant from Ministry of Research and Technology:    
"Bilateral Cooperation between Hellenic Republic and Georgia".

\vfill

\begin{thebibliography}{99}
 
\bibitem{weingarten}D.Weingarten. Nucl.Phys.B210 (1982) 229 \\
 A.Maritan and C.Omero. Phys.Lett. B109 (1982) 51\\
 T.Sterling and J.Greensite. Phys.Lett. B121 (1983) 345 \\
 B.Durhuus,J.Fr\"ohlich and T.Jonsson. Nucl.Phys.B225 (1983) 183 \\
 J.Ambj\o rn,B.Durhuus,J.Fr\"ohlich and T.Jonsson. Nucl.Phys.B290 (1987) 480\\
 T.Hofs\"ass and H.Kleinert. Phys.Lett. A102 (1984) 420 \\
 M.Karowski and H.J.Thun. Phys.Rev.Lett. 54 (1985) 2556\\
 F.David. Europhys.Lett. 9 (1989) 575 

\bibitem{gross}D.Gross. Phys.Lett. B138 (1984) 185\\
V.A.Kazakov. Phys.Lett. B150 (1985) 282\\ 
F.David.  Nucl.Phys. B257 (1985) 45\\
J.Ambj\o rn, B.Durhuus and J.Fr\"ohlich. Nucl.Phys. B257 (1985) 433\\
V.A.Kazakov, I.K.Kostov and A.A.Migdal. Phys.Lett. B157 (1985) 295


\bibitem{ambjorn}J.Ambjorn, B.Durhuus and T.Jonsson. Quantum geometry.
Cambridge Monographs on Mathematical Physics. Cambridge 1998;  
J.Phys.A 21 (1988) 981

\bibitem{helfrich} W.Helfrich. Z.Naturforsch. C28 (1973) 693; J.Phys.(Paris) 46 (1985) 1263\\
L.Peliti and S.Leibler. Phys.Rev.Lett. 54 (1985) 1690\\
A.Polykov. Nucl.Phys.B268 (1986) 406\\
D.Forster. Phys.Lett. 114A (1986) 115\\
H.Kleinert. Phys.Lett. 174B (1986) 335\\
T.L.Curtright and et.al. Phys.Rev.Lett. 57 (1986)799; Phys.Rev. D34 (1986) 3811\\
F.David. Europhys.Lett. 2 (1986) 577\\
P.O.Mazur and V.P.Nair. Nucl.Phys. B284 (1987) 146\\
E.Braaten and C.K.Zachos. Pys.Rev. D35 (1987) 1512\\
E.Braaten, R.D.Pisarski and S.M.Tye. Phys.Rev.Lett. 58 (1987) 93\\
P.Olesen and S.K.Yang. Nucl.Phys. B283 (1987) 73\\
R.D.Pisarski. Phys.Rev.Lett. 58 (1987) 1300


\bibitem{amb}R.V. Ambartzumian, G.K. Savvidy , K.G. Savvidy\\
and G.S. Sukiasian. Phys. Lett. B275 (1992) 99\\
G.K. Savvidy and K.G. Savvidy. Mod.Phys.Lett. A8 (1993) 2963\\
G.K. Savvidy and K.G. Savvidy. Int. J. Mod. Phys. A8 (1993) 3993\\
G.K.Savvidy and F.J.Wegner. Nucl.Phys.B413(1994)605\\
G.K. Savvidy and K.G. Savvidy. Phys.Lett. B324 (1994) 72


\bibitem{sav2}G.K. Savvidy and K.G. Savvidy. 
Phys.Lett. B337 (1994) 333;\\ Mod.Phys.Lett. A11 (1996) 1379.\\
G.K. Savvidy, K.G. Savvidy and P.K.Savvidy  Phys.Lett. A221 (1996) 233

\bibitem{kramers}H.A.Kramers and G.H.Wannier. Phys.Rev. 60 (1941) 252\\
L.Onsager. Phys.Rev. 65 (1944) 117\\
M.Kac and J.C.Ward. Phys.Rev. 88 (1952) 1332\\
C.A.Hurst and H.S.Green. J.Chem.Phys. 33 (1960) 1059


\bibitem{george}T.Jonsson and G.K.Savvidy. Phys.Lett.B449 (1999) 254\\ 
T.Jonsson and G.K.Savvidy. Nucl.Phys. B575 (2000) 661\\
G.K.Savvidy. JHEP 0009 (2000) 044

\bibitem{nigh} M.P.Nightingale. Physica 83A (1976) 561

\bibitem{baxter} R.J.Baxter, Exactly solved models in statistical
mechanics.\\ Academic Press, London 1982

\bibitem{zamolodchikov} A.B. Zamolodchikov. Commun.Math.Phys.79 (1981) 489\\ 
V.V.Bazhanov and R.J.Baxter. J.Stat.Phys. 69(1992) 453

\bibitem{polyakov}A.Polyakov. Gauge fields and String. 
(Harwood Academic Publishers, 1987)\\
E.Fradkin, M.Srednicky and L.Susskind. 
Phys.Rev. D21 (1980) 2885\\
C.Itzykson. Nucl.Phys. B210 (1982) 477 \\
A.Casher, D.Foerster and P.Windey. 
Nucl.Phys. B251 (1985) 29 

\bibitem{pie}R. Pietig and F.J. Wegner. Nucl.Phys. B466 (1996) 513;\\
Nucl.Phys.B525 (1998) 549

\bibitem{bathas}G.K.Bathas et.al. Mod.Phys.Lett. A10 (1995) 2695\\
D.Johnson and R.K.P.C.Malmini,  Phys.Lett. B378 (1996) 87\\
M.Baig, D.Espriu, D.Johnson and R.K.P.C.Malmini,\\ 
J.Phys. A30 (1997) 407; J.Phys. A30 (1997) 7695\\
G.Koutsoumbas et.al. Phys.Lett. B410 (1997) 241

\bibitem{cappi}A.Cappi, P.Colangelo, G.Gonella and A.Maritan, Nucl. Phys. B370 (1992) 659\\
G.Gonnella, S.Lise and A.Maritan, Europhys. Lett. 32 (1995) 735\\
E.N.M.Cirillo and G.Gonella. J.Phys.A: Math.Gen.28 (1995) 867\\



\end{thebibliography}
\end{document}